\documentclass[lettersize,journal]{IEEEtran}
\usepackage{amsmath,amsfonts}
\usepackage{algorithmic}
\usepackage{algorithm}
\usepackage{array}
\usepackage[caption=false,font=normalsize,labelfont=sf,textfont=sf]{subfig}
\usepackage{textcomp,xcolor,textpos}
\usepackage{stfloats}
\usepackage{url}
\usepackage{verbatim}
\usepackage{graphicx}
\usepackage{float}
\newfloat{footnote}{hb}
\graphicspath{{images_eps/},{.}}
\DeclareGraphicsExtensions{.pdf,.jpeg,.png,.jpg}
\usepackage{cite}
\usepackage{multirow,multicol}

\begin{document}
%\begin{textblock*}{\textwidth}(1cm,1cm)
%\end{textblock*}
%\title{Experimental Demonstration of CMOS-Neuronal and HfO$_x$ OxRAM-Synaptic Building Blocks for BNN/TNN}
\title{Time-multiplexed In-memory computation scheme for mapping Quantized Neural Networks on hybrid CMOS-OxRAM building blocks}
\author{Sandeep Kaur Kingra$^{*}$, Vivek Parmar$^{*}$, Manoj Sharma and Manan Suri\vspace{-3em}
\thanks{(c) Copyright 2022 IEEE. This work was supported in part by SERB-CRG/2018/001901 and IITD-FIRP grant. S. K. Kingra, V. Parmar, M. Sharma and M. Suri are with the Indian Institute of Technology Delhi, New Delhi.} 
\thanks{$^*$S. K. Kingra and V. Parmar contributed equally for the manuscript.}
\thanks{Corresponding Authors: Manan Suri (manansuri@ee.iitd.ac.in).
}}

\markboth{IEEE Transactions on Nanotechnology,~Vol.~x, No.~x, Month~2022}%
{Kingra \MakeLowercase{\textit{et al.}}: Time-multiplexed IMC Scheme for mapping QNN}


\maketitle

\begin{abstract}
In this work, we experimentally demonstrate two key building blocks for realizing Binary/Ternary Neural Networks (BNNs/TNNs): (i) 130 nm CMOS based sigmoidal neurons and (ii) HfO$_2$ based multi-level (MLC) OxRAM-synaptic blocks. An optimized vector matrix multiplication (VMM) programming scheme that utilizes the two building blocks is also presented. Compared to prior approaches that utilize differential synaptic structures, a single device per synapse with two sets of READ operations is used. Proposed hardware mapping strategy shows performance change of $<$5\% (decrease of 2-5\% for TNN, increase of 0.2\% for BNN) compared to software-based implementation with significant memory savings in the order of 16-32$\times$ for classification problem on Fashion MNIST (FMNIST) dataset. Impact of OxRAM device variability on the performance of Hardware BNN/TNN is also analyzed.  
\end{abstract}


\begin{footnote*}[!b]
© 2022 IEEE.  Personal use of this material is permitted.  Permission from IEEE must be obtained for all other uses, in any current or future media, including reprinting/republishing this material for advertising or promotional purposes, creating new collective works, for resale or redistribution to servers or lists, or reuse of any copyrighted component of this work in other works.
\end{footnote*}
\section{Introduction}
With advent of artificial intelligence (AI), optimizations of compute architectures to reduce energy and latency have received significant attention\cite{Sze_2017}. CMOS-based von-Neumann architectures suffer limitations in terms of leakage energy (with scaling) as well as memory bottlenecks\cite{KwonR18}. To address the same, novel concepts such as in-memory computing (IMC) \cite{Sebastian_2020} as well as quantized low-precision computation\cite{HubaraCSEB16} have been successfully demonstrated for AI applications. Recent research in quantized neural networks (QNN) focusing on low-precision compute i.e. Binary/Ternary Neural Networks (BNN/TNN) \cite{HubaraCSEB16,DengJPWL18,YinOYLLLW18,Yins_2020} has created an opportunity to realize computational benefits by exploiting inherent analog computing capabilities of emerging resistive memories \cite{LaborieuxBHKNVP21,Esmanhotto20}. Some of the NVM (non-volatile memory) technologies explored for IMC applications for analog multiplication include: Flash\cite{nvcim,Choi_2020}, RRAM (resistive random access memory)\cite{Bocquet_2018,Huang_2019,Yin_2020,Sebastian_2020,Kingra_2020}, and MRAM (magnetoresistive RAM)\cite{Chang_2019vlsi,Gao_2020,Elbtity_2021,misba_arxiv}. RRAM based XNOR bitcells provide following advantages: (i) less area and non-volatility compared to SRAM ($\approx$150 F$^2$ per bitcell), (ii) lower operating voltages and faster memory access time compared to Flash, and (iii) lower fabrication cost, reduced area and write energy compared to MRAM.

\begin{figure}[tb]
  \centering
  \includegraphics[width=\linewidth]{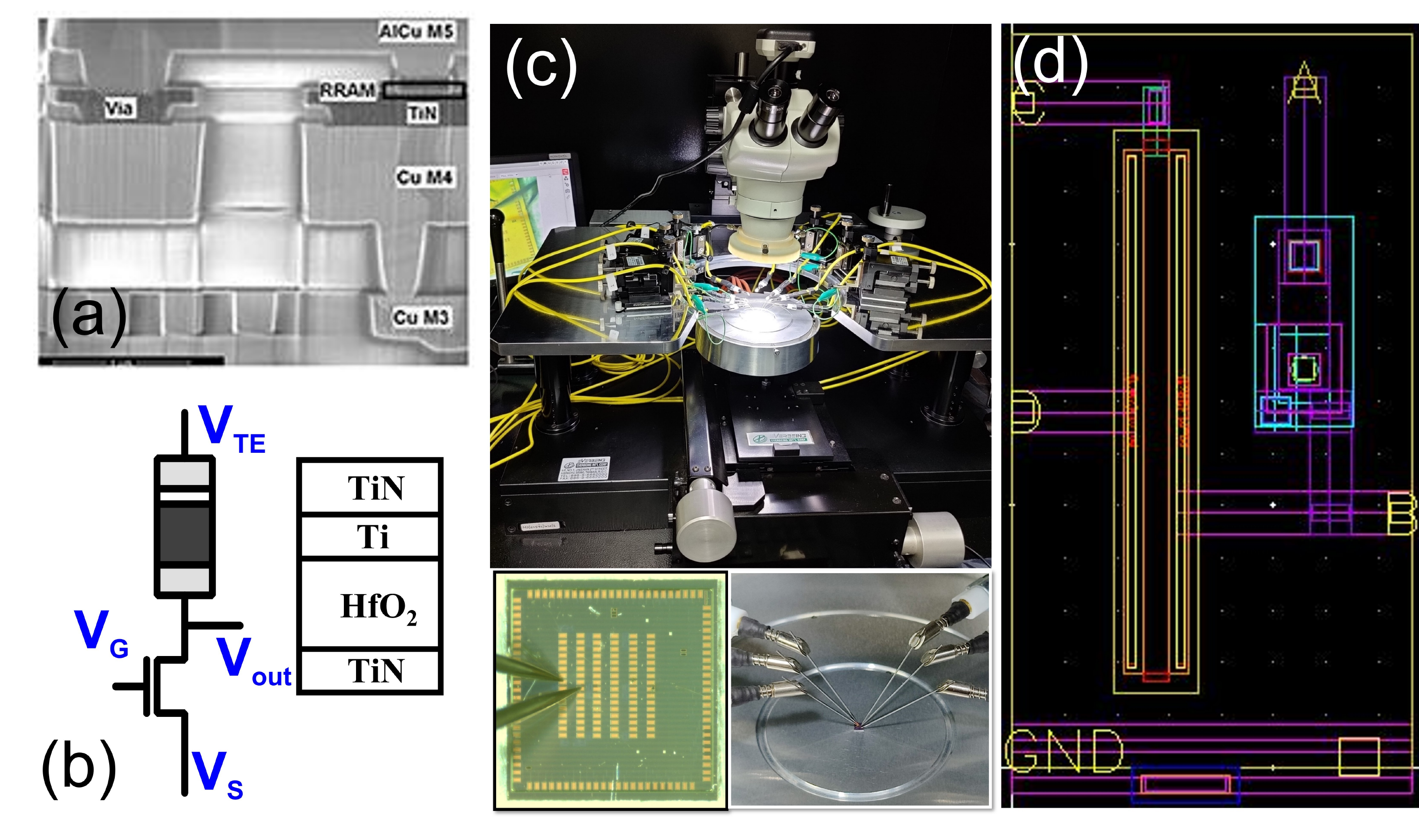}
  \caption{(a) TEM cross section of the integrated $TiN/HfO_2/Ti/TiN$ OxRAM\cite{Parmar_2022}. (b) 1T-1R bitcell with OxRAM device stack. (c) Test setup used in this study. (d) Layout of the fabricated 1T-1R bitcell using 130 nm CMOS technology.}
  \label{Fig1}
\end{figure}

BNNs assume only binary values (`+1', `-1') while TNNs assume ternary values (`+1', `0', `-1') for both synaptic weights and neuronal activations. BNN based on differential synapses have been explored {demonstrating error resilience to programming failures\cite{Bocquet_2018,Yin_2020}}. Recent studies have also shown TNNs realized by using similar BNN hardware that offer sparsity benefits \cite{YinOYLLLW18,Yins_2020,LaborieuxBHKNVP21}. {Recently some groups have investigated incorporation of device non-idealities during training to further improve error resilience of the final implementation\cite{misba_arxiv}. However, this requires increased computational resource during training to account accurately for hardware variability. Training using conventional algorithms while resulting in less accuracy may still result in tolerable error resilience at lower computational costs.} In this study, we propose BNN/TNN implementation using fabricated 1T-1R device-based synapses (Fig.~\ref{Fig1}(a)) by exploiting MLC \cite{Hsieh19} OxRAM programming and CMOS sigmoid neuron (on-chip) for classification on Fashion MNIST (FMNIST) dataset \cite{FMNIST1}. Key contributions of the work are listed below: 

\begin{enumerate}
    \item Demonstration of a batchnorm-free end-to-end implementation (synapse and neuron on same die) of CMOS-OxRAM based hardware QNNs (BNN/TNN).
    \item Time-multiplexed VMM computation using single device per synapse and sequentially-applied input activation.
    \item Performance comparison of QNNs based on BNN/TNN as well as impact of MLC state distributions using FMNIST dataset. 
    \item Proposed hardware mapping strategy for QNN shows performance change of $<$5\% (decrease of 2-5\% for TNN, increase of 0.2\% for BNN) compared to software-based implementation.
\end{enumerate}

The manuscript is organized as follows: Section~\ref{lit} describes key concepts such as BNN/TNN along with hardware implementations, Section~\ref{oxram} presents details of fabricated circuits (sigmoid neuron and 1T-1R synapse) and device characterization results, Section~\ref{vmm} describes proposed VMM computation scheme for BNN/TNN realization, Section~\ref{result} describes key results based on network simulations and finally Section~\ref{conc} provides concluding remarks.

%demonstration of a batchnorm-free end-to-end implementation of OxRAM based hardware QNNs (BNN/TNN) utilizing single device per synapse while demonstrating a marginal performance change of $<$2\% ($\downarrow$2\% for TNN, $\uparrow$0.2\% for BNN) compared to ideal software QNN. 

\begin{figure}[tb]
  \centering
  \includegraphics[width=0.8\linewidth]{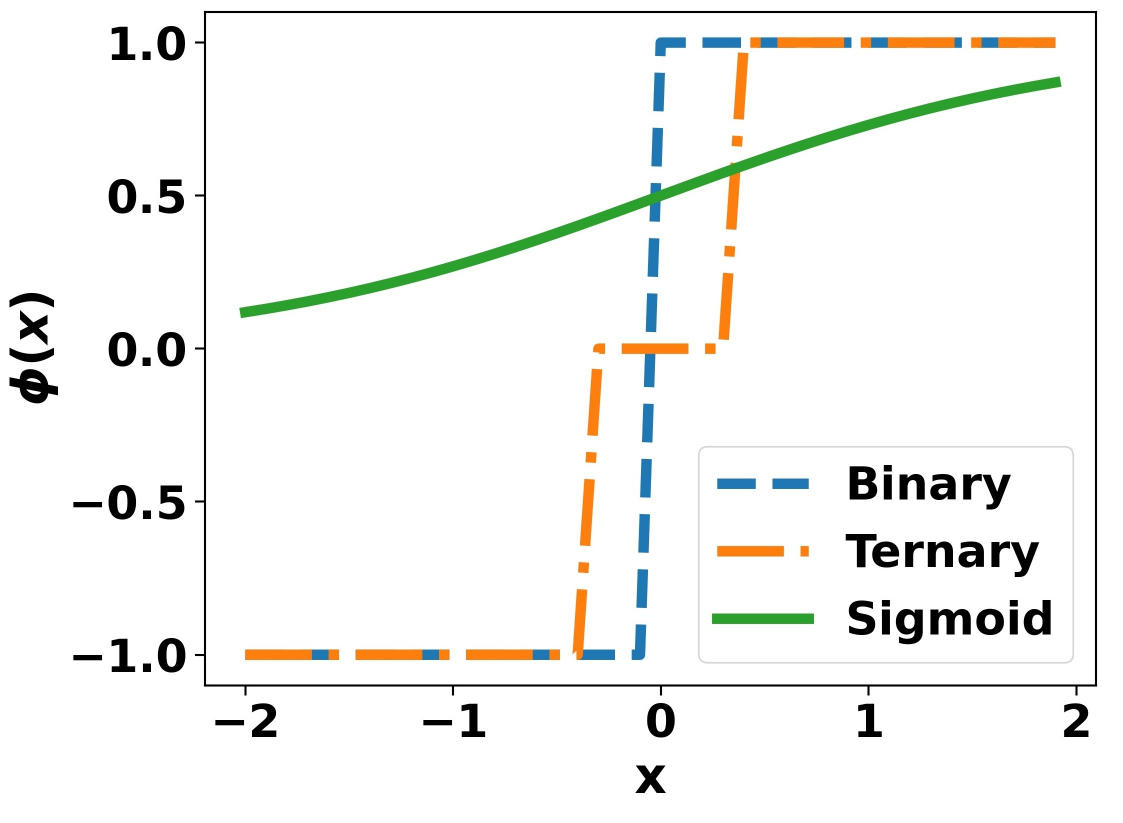}
  \caption{Comparison of responses of activation functions used in the study.}
  \label{Fig1a}
\end{figure}
\section{Prior Art}
\label{lit}
QNNs enable an efficient mapping of neural network computations to analog hardware by restricting precision\cite{Sze_2017} while simultaneously offering significant resilience to errors\cite{Parmar_2021}. Extreme forms of QNN investigated in literature include BNN (2 states/ 1-bit) and TNN (3 states/ 2-bit). BNNs are realized in digital hardware in form of XNOR gates followed by popcount circuits. TNNs add `0' state and replace XNOR with Gated-XNOR\cite{DengJPWL18}. Gating is realized by checking non-zero status of both inputs thus avoiding unnecessary computation of XNOR operations. This further offers benefits in terms of sparsity (for storage) as well as pruning (for computations) in comparison to BNN. The key difference exists between both in terms of activation functions employed as shown below in Eq.~(\ref{eq1}) and Eq.~(\ref{eq2}).
\begin{equation}
    Act_{binary}(x) =
    \begin{cases}
     +1, & \text{$x\geq0$}\\
     -1, & \text{$x<0$}
    \end{cases}
    \label{eq1}
\end{equation}
\begin{equation}
    Act_{ternary}(x) =
    \begin{cases}
     +1, & \text{$x>r$}\\
     -1, & \text{$x<-r$}\\
     0, & \text{$x \leq abs(r)$}
    \end{cases}
    \label{eq2}
\end{equation}    

BNN hardware implementations based on In-Memory Computing (IMC) typically employ usage of differential weights and inputs to realize dual polarity (`+1',`-1') in hardware\cite{Bocquet_2018,Yin_2020,Yins_2020}. Recent studies have also shown TNNs realized by using similar hardware that offer sparsity benefits \cite{YinOYLLLW18,Yins_2020,LaborieuxBHKNVP21} by employing a 3$^{rd}$ state combination for the synapse (both devices in HRS) and activations (all 0's). This leads to savings in energy dissipation for computation. {Some studies have explored benefits of employing sigmoid neurons with binarized weights to improve accuracy\cite{Elbtity_2021}. However, this can potentially lead to implementation complexities since the resulting output may not conform to the signal requirement of BNN/TNN as sigmoid can result in an analog value (see Fig.~\ref{Fig1a}). To perform computation for all layers directly based on output of previous layer it becomes essential to use activation functions described in Eq.~(\ref{eq1}) and Eq.~(\ref{eq2}) for all the intermediate layers.}
\begin{figure}[tb]
  \centering
  \includegraphics[width=\linewidth]{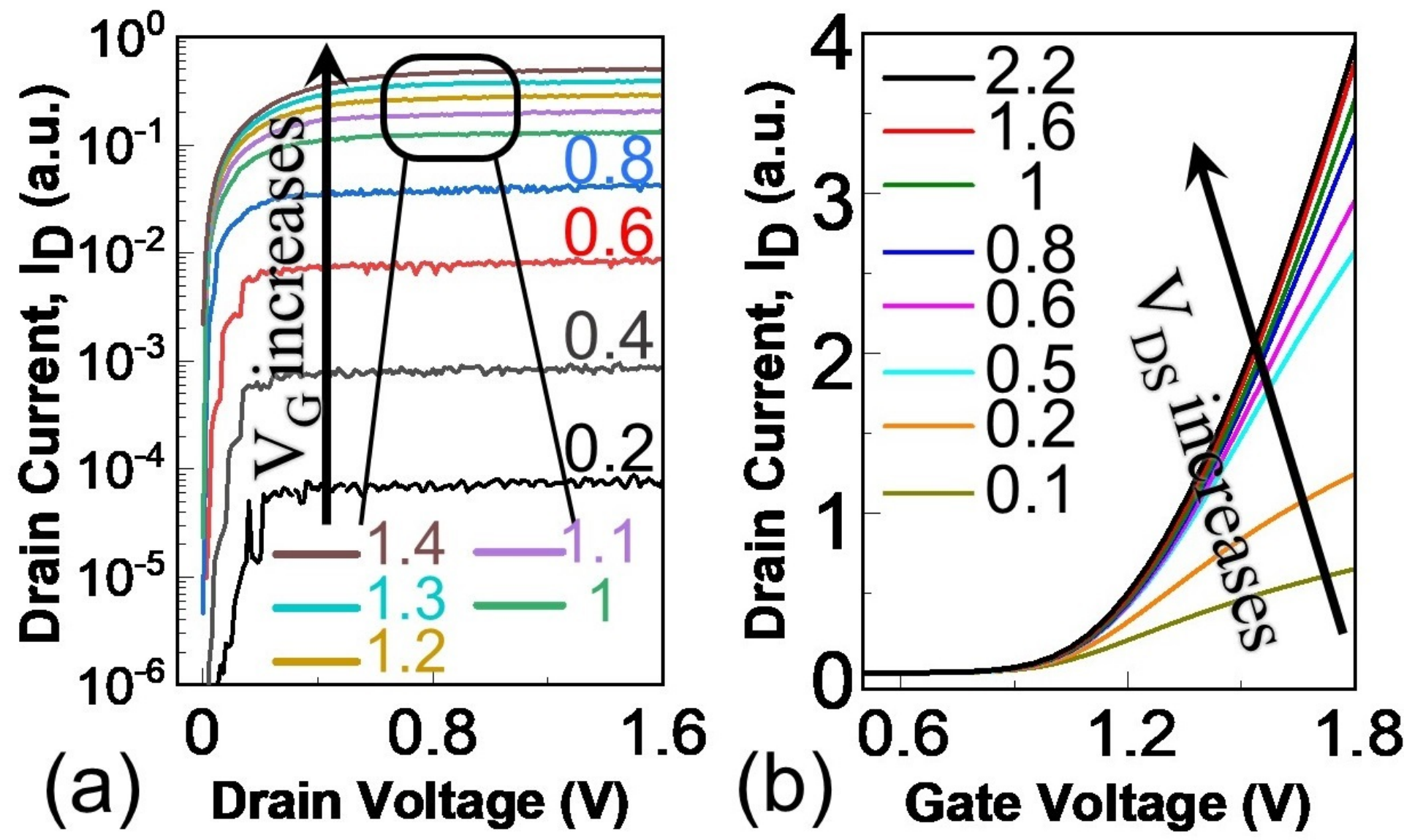}
  \caption{Measured NMOS selector device characteristics: (a) I$_D$ vs V$_{DS}$, and (b) I$_D$ vs V$_{GS}$.}
  \label{Fig2}
\end{figure}

\begin{figure}[tb]
  \centering
  \includegraphics[width=\linewidth]{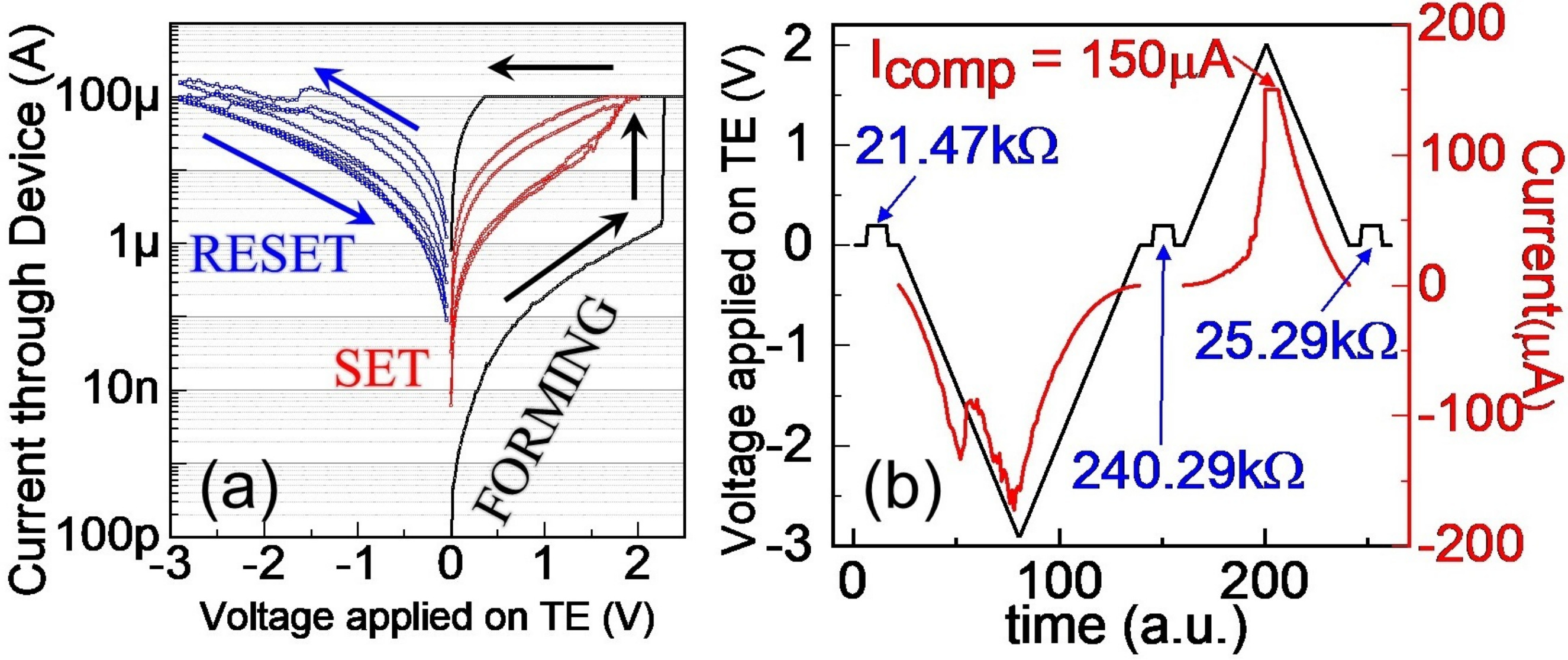}
  \caption{(a) DC-IV characteristics of 1T-1R device highlighting Forming, SET and RESET switching action. (b) Device switching action observed during pulse test. $V_{read}$ = 0.2 V, $V_{gate}$ = 1.2 V, is used to read the OxRAM resistance state.}
  \label{Fig3}
\end{figure}

\begin{figure}[t]
  \centering
  \includegraphics[width=0.9\linewidth]{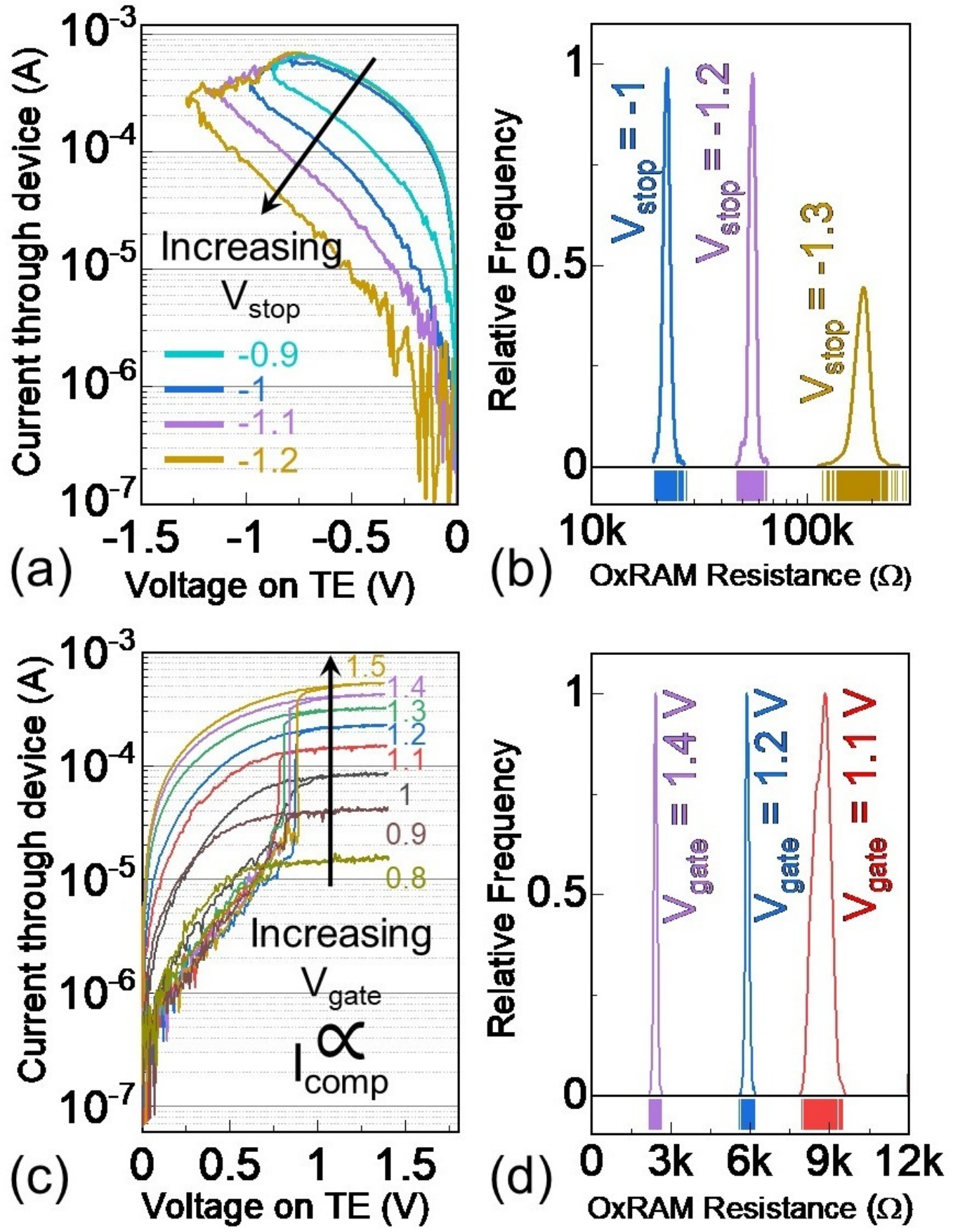}
  \caption{MLC characteristics exhibited by OxRAM device by varying: (a) $V_{stop}$ during RESET programming and gaussian distribution of 3 resistive states selected in the HRS region (with C2C/D2D distribution) for realizing BNNs/TNNs in (b). (c) $V_{gate}$ to tune $I_{comp}$ during SET programming and gaussian distribution of 3 resistive states (with C2C/D2D distribution) selected from possible states shown in (c) for realizing BNNs/TNNs in (d).}
  \label{Fig4}
\end{figure}
\begin{figure}[htb]
  \centering
  \includegraphics[width=0.95\linewidth]{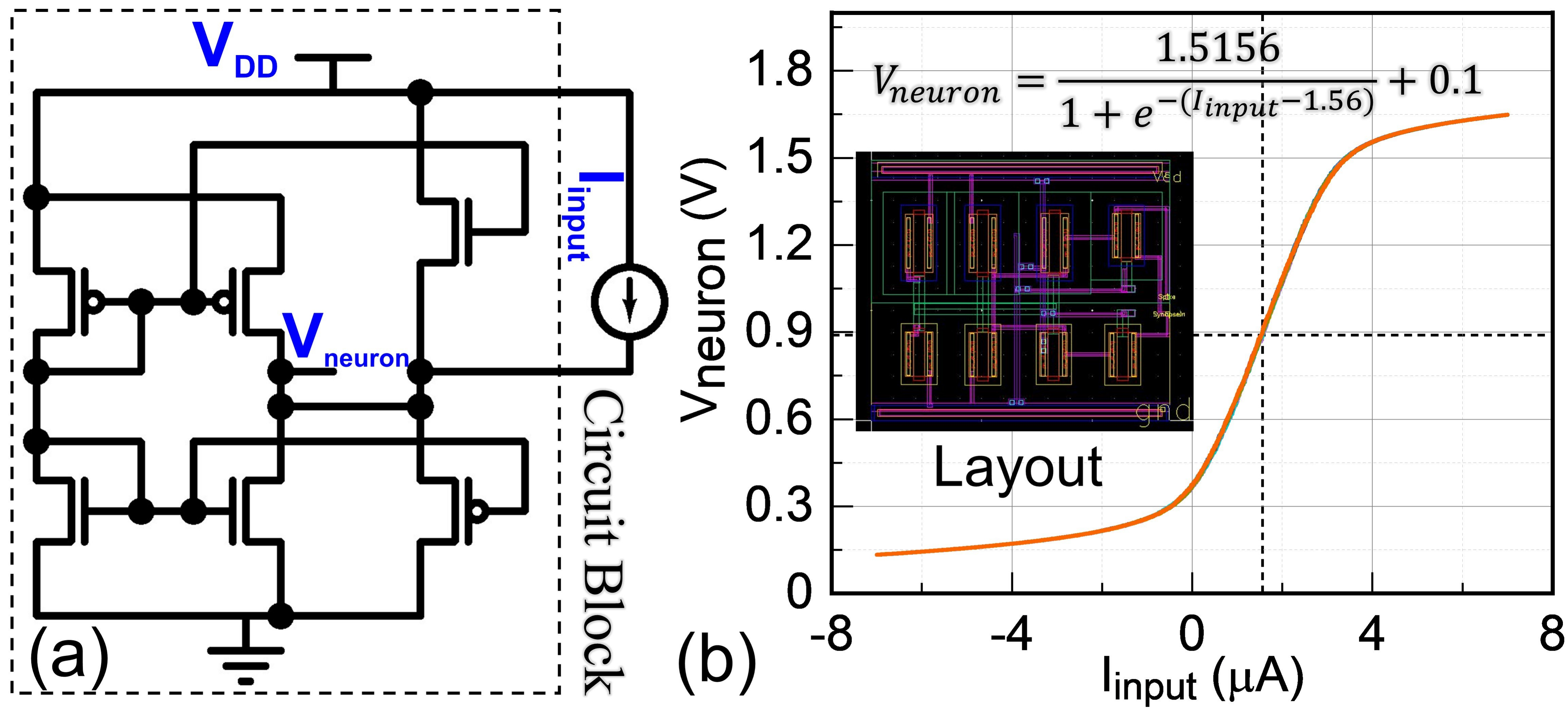}
  \caption{(a) Circuit schematic of 6T CMOS sigmoid neuron. (b) Measured sigmoid transfer curve validating circuit functionality along with its layout (inset).}
  \label{Fig5}
\end{figure}

\section{Fabricated CMOS-OxRAM circuits}
\label{oxram}
A single resistive memory cell comprises of 1T-1R (1 transistor-1 resistor), where NMOS transistor acts as the selection element in the array. TiN/HfO$_2$/Ti/TiN OxRAM devices (referred as R in 1T-1R), having 300 nm cross point diameter are monolithically integrated on top of 130 nm CMOS baseline process (see Fig.~\ref{Fig1}(a)). {The basic cell structure and device stack are shown in Fig.~\ref{Fig1}(b).} The experimental test setup used in this study is shown in Fig.~\ref{Fig1}(c). In Fig.~\ref{Fig1}(d), the layout of 1T-1R bitcell is shown highlighting the OxRAM device layers. Experimentally measured NMOS selector characteristics (I$_D$ vs V$_{DS}$, I$_D$ vs V$_{GS}$) are shown in Fig.~\ref{Fig2}.  The forming procedure is required on pristine OxRAM devices before executing SET/RESET programming operations. DC-IV curves showing FORMING, SET and RESET switching actions of 1T-1R OxRAM devices are presented in Fig.~\ref{Fig3}(a). Fig.~\ref{Fig3}(b) illustrates pulse switching behavior of the 1T-1R structure. {For measuring the device resistance state, a 10 $\mu$s read pulse with V$_{read}$ = 0.2 V is used.} MLC programming of the 1T-1R structure is achieved (see Fig.~\ref{Fig4}) by modulating $V_{stop}$ for HRS (High Resistance State) and $I_{comp}$ (compliance current) for LRS (Low Resistance State). In Fig.~\ref{Fig4}(a), by sweeping $V_{stop}$ from -0.9 V to -1.2 V (applied at top electrode), different HRS region states are realized. {For MLC in LRS region, $V_{gate}$ of NMOS selector is tuned to control $I_{comp}$. In Fig.~\ref{Fig4}(c), by sweeping the $V_{gate}$ from 0.8 V to 1.5 V, multiple LRS states are realized.} To realize hardware BNN/TNN, we selected three distinct resistive states in HRS/LRS region. These well separable programming states are selected based on programming reproducibility. Fig.~\ref{Fig4}(b,d) presents the HRS/LRS distribution for selected 1T-1R bitcell states including C2C/D2D (C=cycle, D=device) variability. As observed, the states are well separable, thereby offering better reliability for BNN/TNN realization.

Using the same baseline 130 nm CMOS we designed and fabricated 6T-sigmoid neurons (Fig.~\ref{Fig5}). {Fig.~\ref{Fig5}(a) presents the circuit schematic for the CMOS sigmoid neurons that acts as current-voltage conversion circuit. Here, $I_{input}$ current acts the input signal to the circuit and output is realized as $V_{neuron}$. Ideal, sigmoid function has output as defined in Eq. (\ref{eq3}),}
\begin{equation}
    Sigmoid(x) = \frac{1}{1+e^{-x}} \label{eq3}
\end{equation}
{
Experimentally characterized sigmoid neuron's transfer function characteristics are shown in Fig.~\ref{Fig5}(b). Here a slight offset in the $V_{neuron}$ output is observed and it is represented as Eq. (\ref{eq4}) where $I_{input}$ is specified in $\mu$A. The layout of the sigmoid neuron is shown as inset in Fig.~\ref{Fig5}(b).}

\begin{equation}
    V_{neuron}(I_{input}) = \frac{1.5156}{1+e^{-(I_{input}-1.56)}} + 0.1 \label{eq4}
\end{equation}

\section{VMM operation mapping}
\label{vmm}
\begin{figure}[tb]
  \centering
  \includegraphics[width=0.8\linewidth]{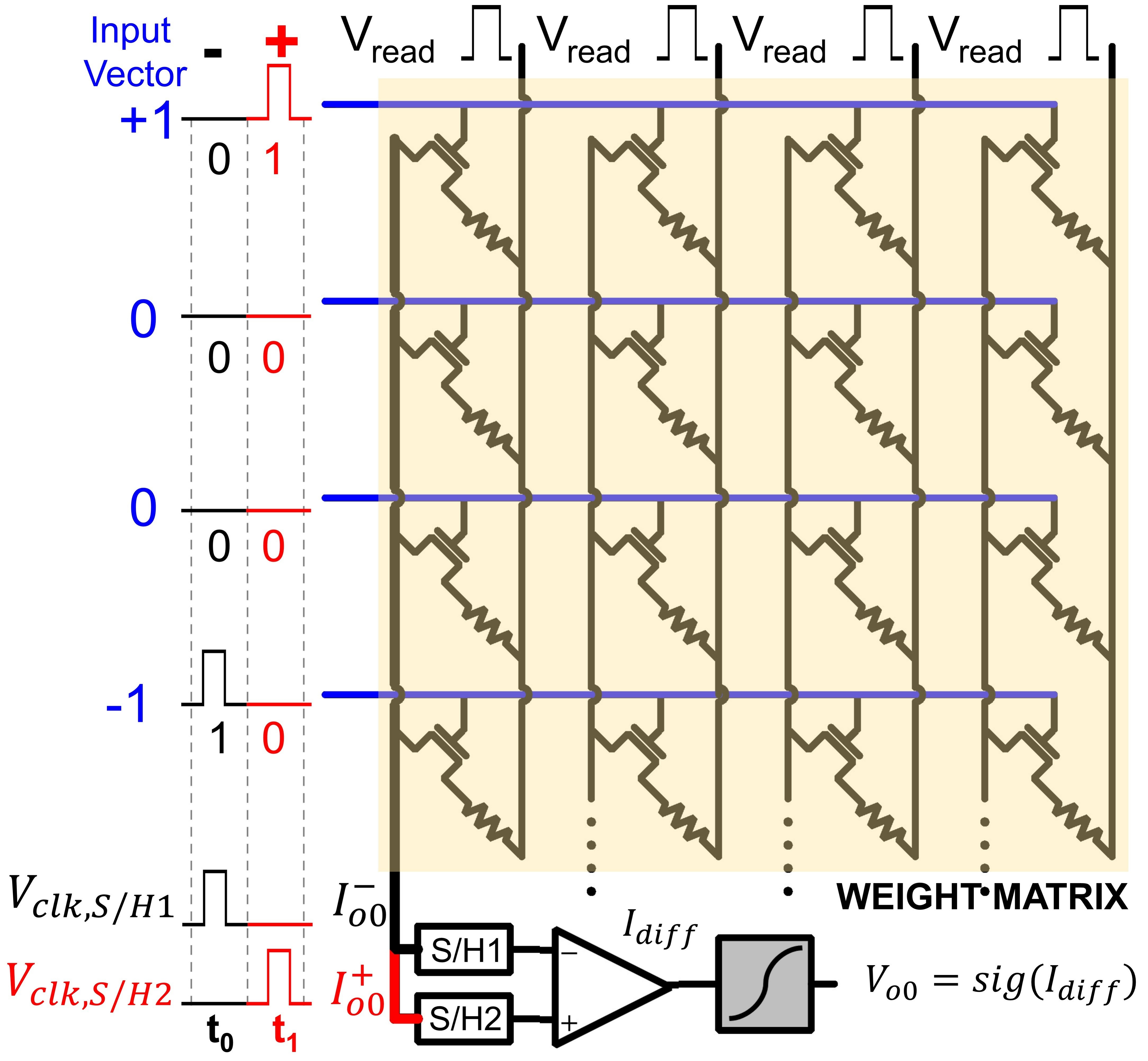}
  \caption{Vector multiplication for BNN/TNN using MLC OxRAM states and 2-step READ operations for applied input vector = [+1, 0, 0, -1]. Note sigmoid neuron based activation is used for class detection.}
  \label{Fig6}
\end{figure}
\begin{figure}[tb]
  \centering
  \includegraphics[width=0.8\linewidth]{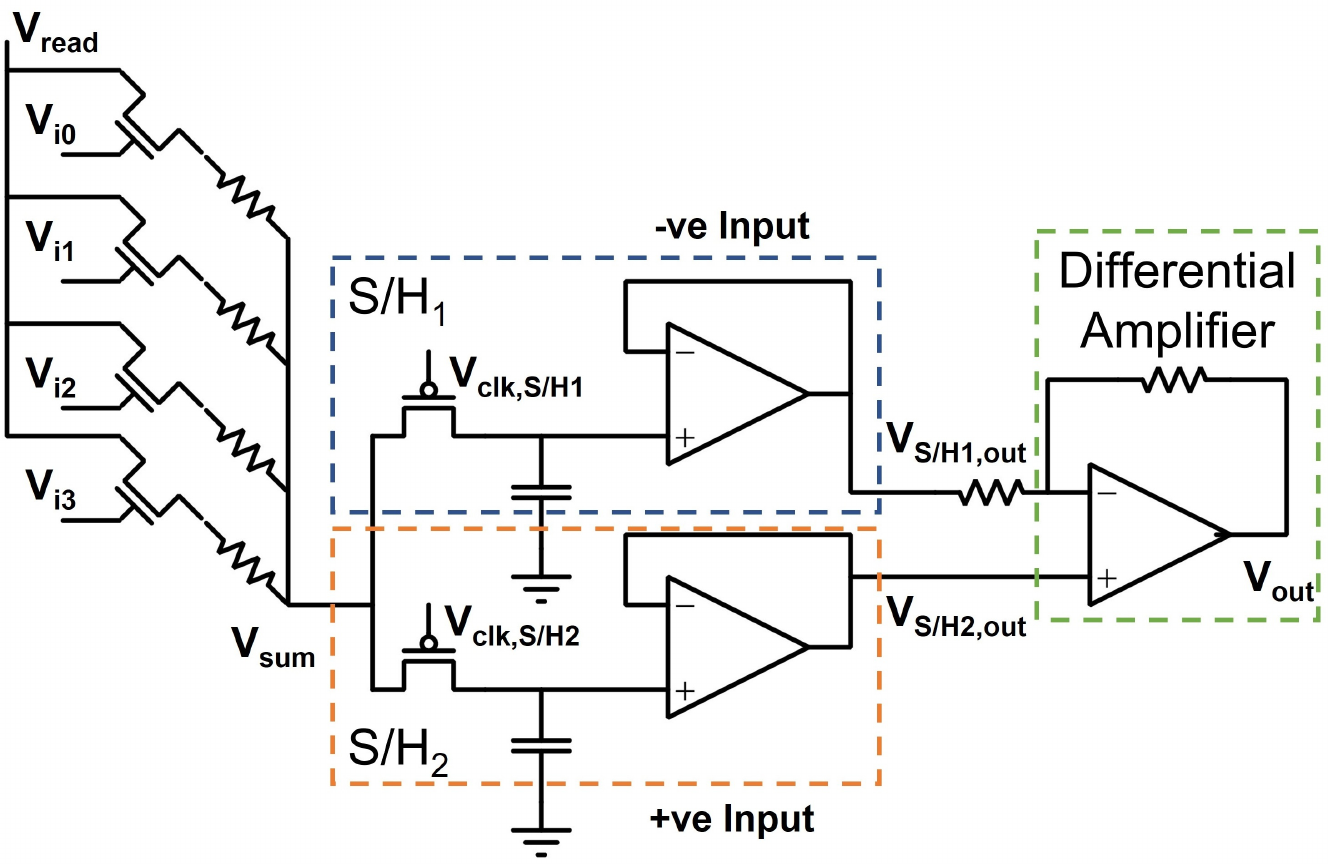}
  \caption{Simulated S/H-based time-multiplexed differentiation circuit for validation of proposed methodology. The circuit is simulated with resistance values based on experimental characterized 1T-1R data along with 130 nm CMOS technology.}
  \label{Fig6b}
\end{figure}
\begin{figure}[tb]
  \centering
  \includegraphics[width=0.95\linewidth]{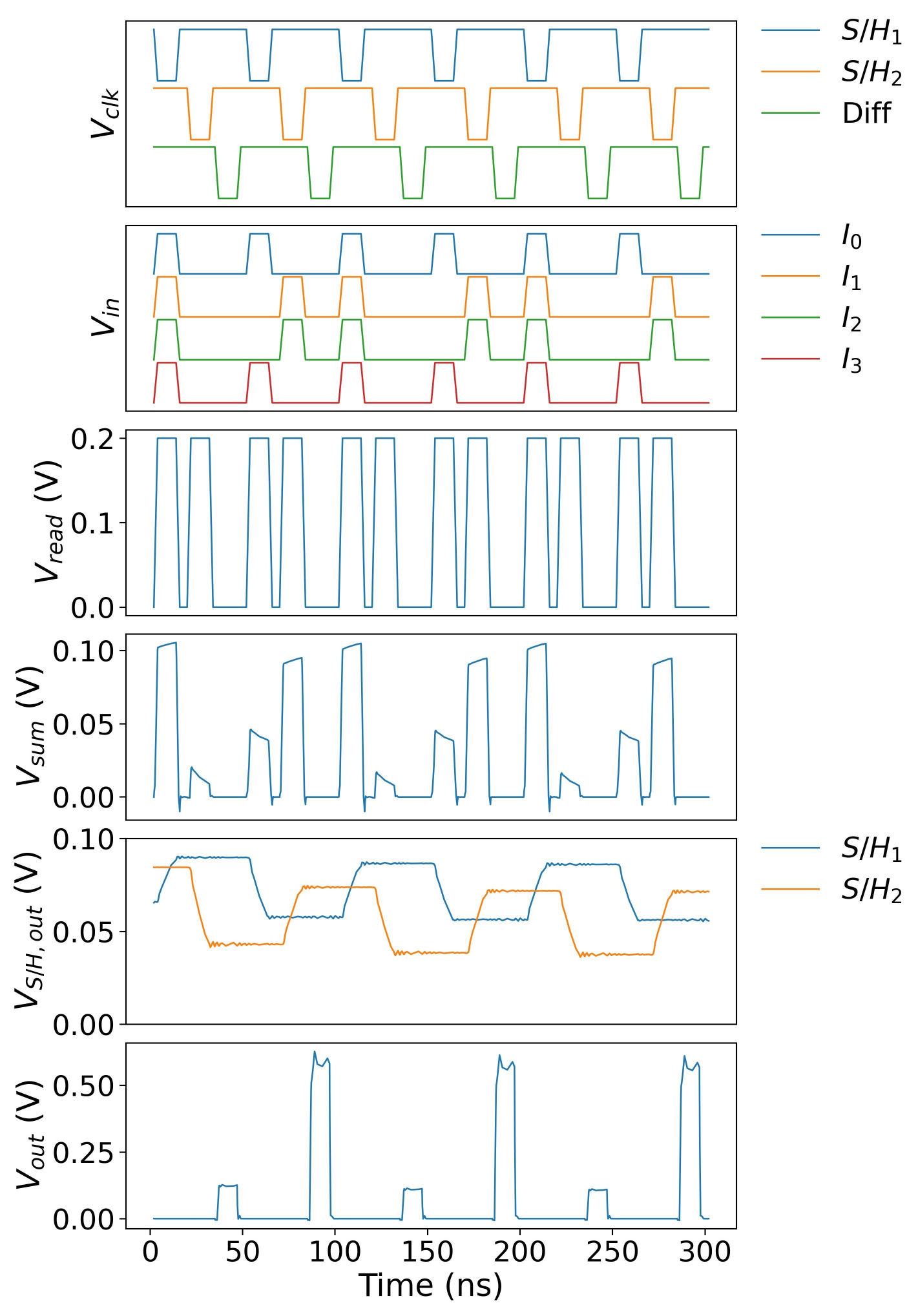}
  \caption{Input and output waveforms based on SPICE simulations of time-multiplexed differential circuit validating proposed methodology. $V_{out}$ is high in case output of $S/H_2$ exceeds output of $S/H_1$. $V_{in}$ and $V_{clk}$ are pulses of amplitude 1.4V. Since clock signals are applied at the gate terminal of PMOS transistor, output is sampled on the input lines when $V_{clk}$ is low. Similarly, final output voltage $V_{out}$ is sampled when $V_{clk,diff}$ is low.}
  \label{Fig6c}
\end{figure}
\begin{figure}[htb]
  \centering
  \includegraphics[width=0.9\linewidth]{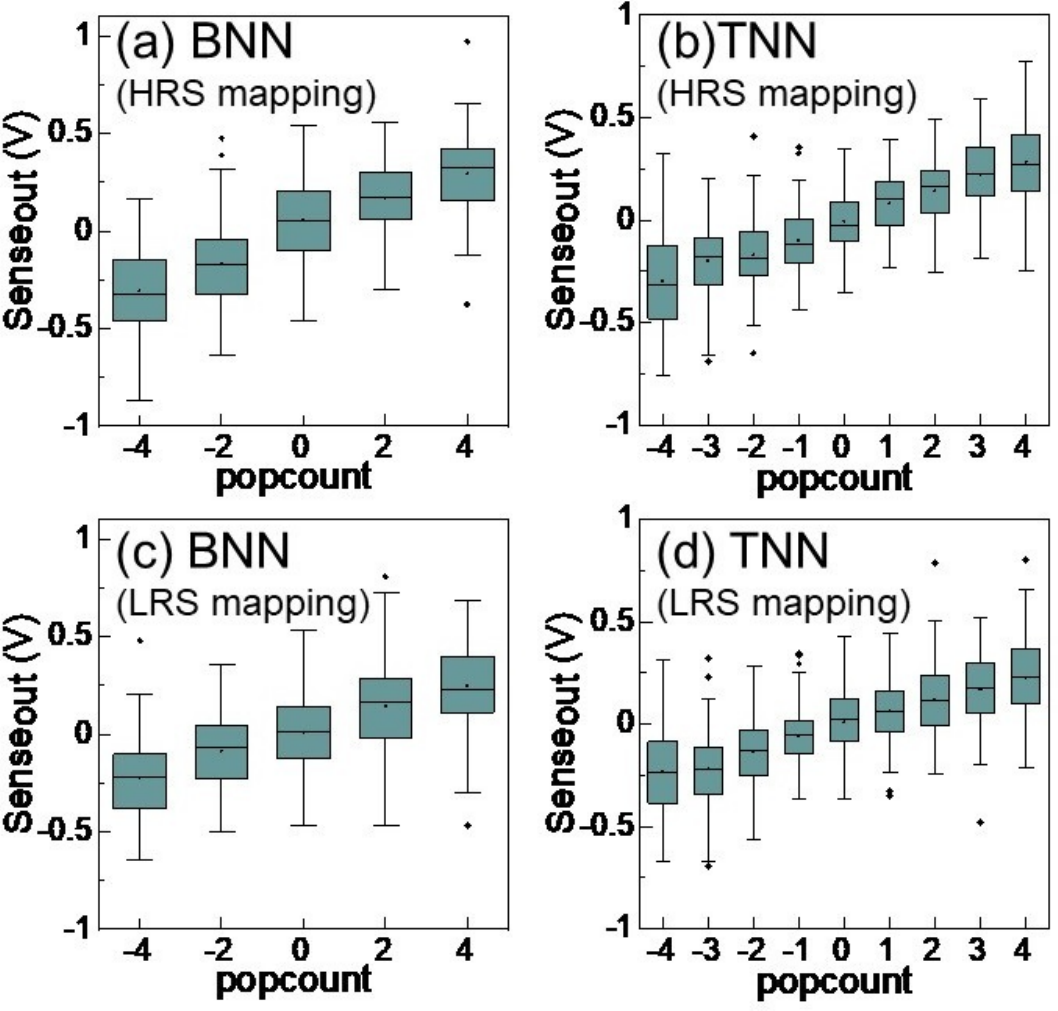}
  \caption{Simulated VMM output voltage for 4$\times$4 matrix of 1T-1R devices using MLC states in LRS region for BNN and TNN in (a),(b) respectively;  similarly using MLC states in HRS region, BNN and TNN Sense$_{out}$ voltages in (c),(d) respectively. Note that x-axis has popcount values possible for BNN/TNN networks.}
  \label{Fig6a}
\end{figure}
In this study, compared to prior work BNN/TNN are realized by exploiting analog MLC states of the OxRAM device and a two-stage input mapping to perform computations.
While implementing analog VMM using resistive memory arrays, output is realized as current integrated along the column \cite{Esmanhotto20}. Sigmoid neuron translates this integrated current ($I_{input}$) into output voltage ($V_{neuron}$). Neuron gain can be fine-tuned by carefully scaling the ($\frac{W}{L}$) of the 6 transistors. For performing vector multiplication, synaptic weights are mapped as OxRAM device conductance (2 states for BNN, 3 states for TNN) while input activations are applied as $V_{gate}$ during two consecutive READ operations at time instances `$t_0$' and `$t_1$' (Fig.~\ref{Fig6}). To map inputs with opposing polarities i.e. (`+1', `-1'), input vector is split into two components i.e. +ve component and –ve component. For +ve component: $V_{gate}$=1.2 V is applied for input $\geq$ 0 (i.e., `+1') and $V_{gate}$=0 V is applied for input $<$ 0 (i.e., `-1'). For –ve component, the vice-versa signals are applied for input `+1' and `-1'. For ternary input value `0', $V_{gate}$=0 V is applied for both input vector components. READ operation for each input vector component generates a column-wise integrated current ($I_{o0}^+$ at $t_0$, $I_{o0}^-$ at $t_1$ for column 0). Since both components are applied at consecutive time instances, Sample \& Hold (S/H) circuits are used to store the integrated current outputs at each time instant. Once both vector components have been applied, S/H outputs are fed to a comparator. The comparator output is then applied to an activation function (here CMOS sigmoid neuron which is used for last layer). 

{To validate the proposed S/H based circuit, extensive circuit simulations are performed. The simulated circuit used for validation is shown in Fig.~\ref{Fig6b}. Corresponding input and output waveforms captured from SPICE simulations are shown in Fig.~\ref{Fig6c}. Input in the form of integrated current ($I_{o0}^+$/$I_{o0}^-$) from a column is first converted to voltage $V_{sum}$ which is captured using phase-delayed clocks ($V_{clk,S/H1}$,$V_{clk,S/H2}$)  by two S/H circuits (S/H$_1$, S/H$_2$). The output voltage ($V_{S/H,out}$) from each S/H circuit is applied at the input of a differential amplifier with gain adjusted based on column length and resistance distribution of the MLC devices. This output can now be fed to any activation function circuit (binary, ternary, sigmoid) to compute the neuron output. As shown in Fig.~\ref{Fig6b}, the proposed S/H-based differential circuit requires 3 op-amp blocks. The area of a single op-amp is estimated to be 45.26 $\mu$m$^2$ thus leading to an overall area cost of $\sim$136 $\mu$m$^2$. Total energy dissipation of the S/H-based differential circuit for a single column operation is estimated to be 83 pJ with a latency of 50 ns. Though there is an increase in the area for the periphery circuits with the proposed approach, the energy cost is not as significant. Prior studies have shown energy cost of programming a single OxRAM device to be $\sim$200-400 pJ \cite{Hirtzlin_2020}. Since the proposed approach enables usage of a single device per synapse, it reduces programming energy cost by half. The area cost of the periphery is already a well-established limitation even with approaches using two devices per synapse that have been addressed using schemes such as multiplexed sensing \cite{Yin_2020}. While the proposed approach shows marginal limitations in terms of area , it still shows significant energy saving in scenarios where weight re-programming is performed at regular intervals ($\approx$2$\times$).} 

{To validate VMM computation methodology, extensive numerical simulations are performed for a $4\times4$ 1T-1R array.} Simulated voltage-output resulting of sigmoid neurons, V$_{neuron}$, (operating in the linear region) from the array is shown in Fig.~\ref{Fig6a}(a,b) for HRS based weight mapping (i.e. the device state mapped on OxRAM array is chosen from Fig.~\ref{Fig4}(b)) and Fig.~\ref{Fig6a}(c,d) for LRS based weight mapping (i.e. the device state mapped on OxRAM array is chosen from Fig.~\ref{Fig4}(d)). Popcount denotes mathematical product of binary/ternary precision vectors utilizing (+1,-1)/(+1,0,-1) as weights and activations. As shown in Fig.~\ref{Fig6a}, TNNs show popcount output at all integer discrete levels (refer  Fig.~\ref{Fig6a}(b,d)). However, BNN discretizes output further to multiples of 2 (see Fig.~\ref{Fig6a}(a,c)). As a result, TNN output is more sensitive to the device variability as compared to BNN.

\section{Network Simulations}
\label{result}
\begin{figure*}[tb]
  \centering
  \includegraphics[width=0.98\linewidth]{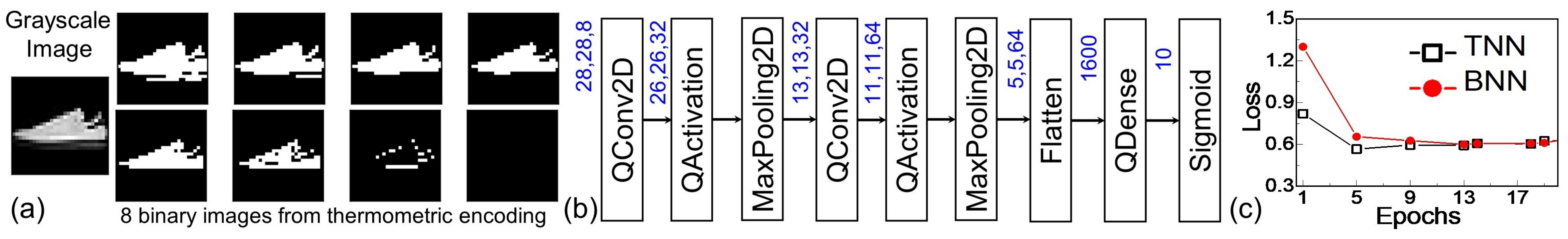}
  \caption{(a) Sample image from FMNIST dataset with binary images resulting from thermometric encoding (channels=8, resolution=32). (b) LeNet-based CNN\cite{LeCun_1989} architecture used in this study. (c) Categorical cross-entropy loss evolution observed during training of BNN/TNN with QKeras\cite{CoelhoKLZNALPPS21} framework.}
  \label{Fig7}
\end{figure*}

To validate the proposed hardware, CNNs with multiple {(binary/ternary)} precisions are trained using QKeras \cite{CoelhoKLZNALPPS21} framework on the FMNIST dataset {based on training algorithms discussed in \cite{Moons_2017,Hwang_2014,TWN1}.} Using thermometric encoding \cite{Rachkovskiy_2005} grayscale images are transformed into 8 channels of binary images (refer Fig.~\ref{Fig7}(a)). Network based on LeNet architecture\cite{LeCun_1989} used for the study is shown in Fig.~\ref{Fig7}(b). Training evolution curves for BNN/TNN (in terms of loss) are shown in Fig.~\ref{Fig7}(c). {Due to batchnorm-free training, optimization converges within few epochs and saturates.}

\begin{comment}
\begin{figure}[t]
  \centering
  \includegraphics[width=0.65\linewidth]{kingra12}
  \caption{Categorical cross-entropy loss evolution observed during training of BNN/TNN with QKeras\cite{CoelhoKLZNALPPS21} framework.}
  \label{Fig7a}
\end{figure}
\end{comment}

\begin{figure}[tb]
  \centering
  \includegraphics[width=0.95\linewidth]{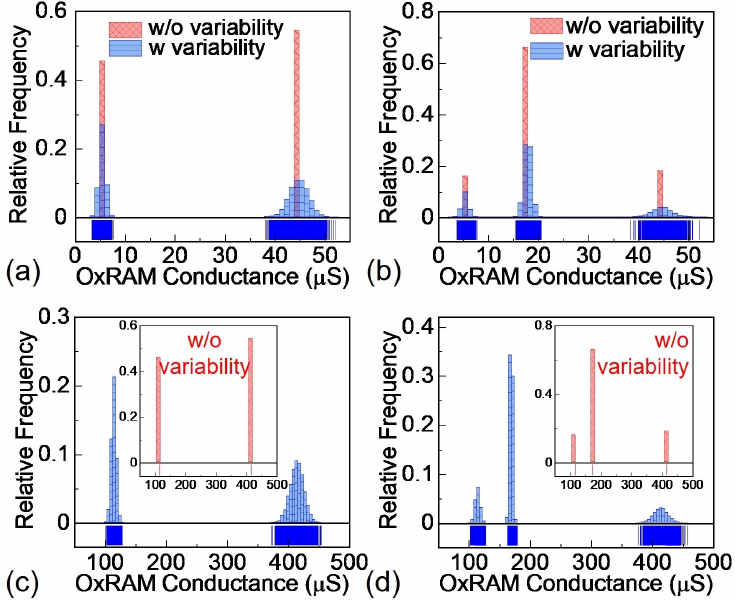}
  \caption{Weight distributions extracted using 3 resistive states from HRS region (as OxRAM device conductance) for simulated BNN and TNN in (a,b) respectively. Similarly, weight distributions extracted using 3 resistive states from LRS region (as OxRAM device conductance) for simulated BNN and TNN in (c,d) respectively.}
  \label{Fig7b}
\end{figure}
\begin{figure}[tb]
  \centering
  \includegraphics[width=0.95\linewidth]{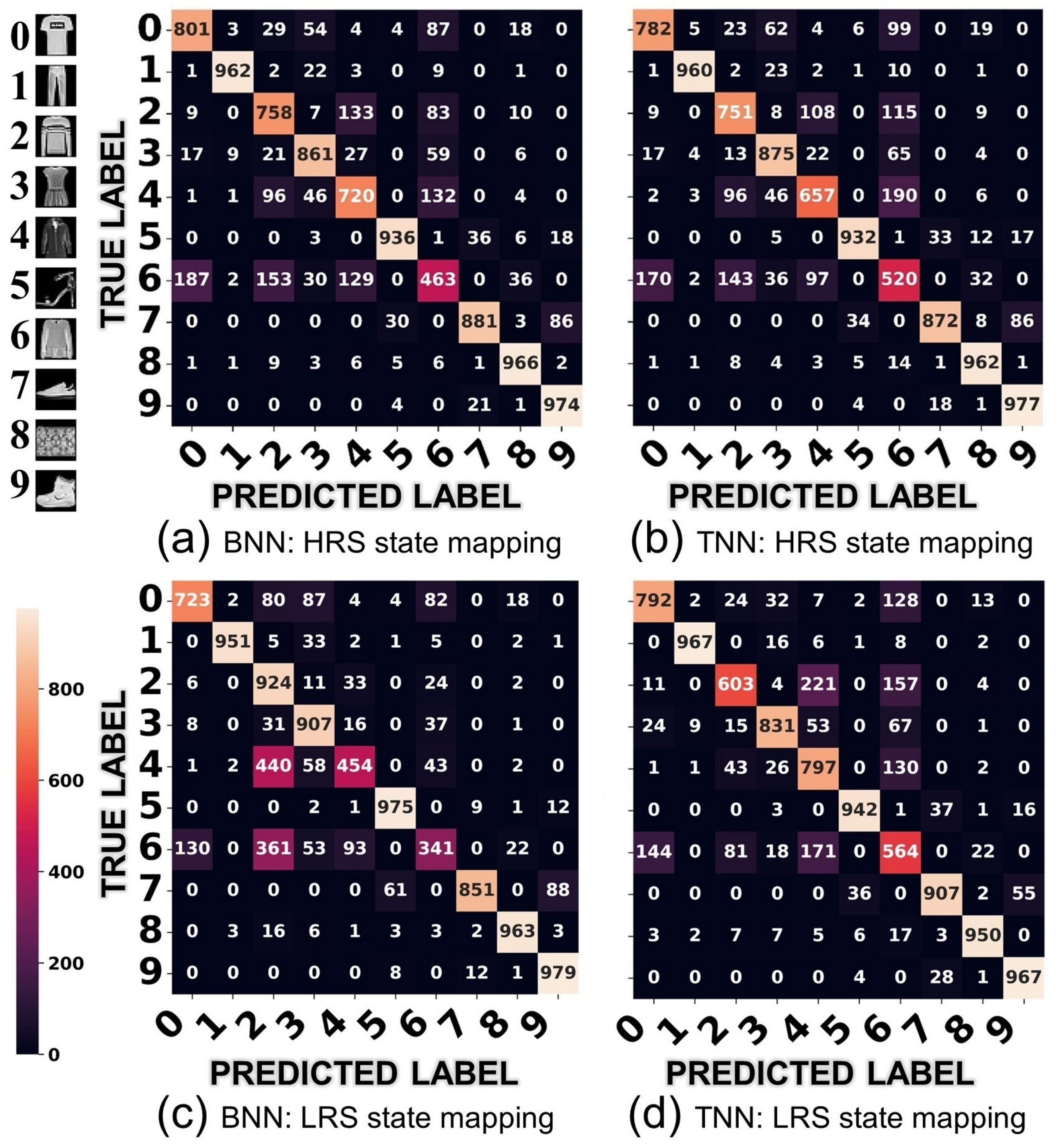}
  \caption{Confusion matrices observed for simulated neural networks including OxRAM device conductance values for BNN/TNN weight realization using selected memory states from HRS region (a,b respectively) and LRS region (c,d respectively).}
  \label{Fig8}
\end{figure}

\begin{table}[tb]
  \centering
  \caption{Performance benchmarking of the proposed BNN/TNN hardware.}
    \begin{tabular}{|c|c|c|c|c|}
    \hline
    \multirow{2}{*}{Platform}& MLC & \multicolumn{3}{c|}{Test Accuracy (\%)} \\
\cline{3-5}          & State & FP32  & TNN   & BNN \\
    \hline
    SW  &  & 91.02 & 85.07 & 82.92 \\
    \hline
    {HW-aware simulation} & LRS Region & -     & 80.63  & 83.2 \\
    \cline{2-5}
    (OxRAM) & HRS Region & -     & 83.2  & 83.1 \\
    \hline
    \end{tabular}%
  \label{tab1}%
\end{table}%

Network simulations incorporating experimentally measured OxRAM device conductance (including C2C/D2D variability) are performed by mapping trained weights to MLC conductance state. Weight distributions for BNN/TNN in terms of HRS device conductance are shown in Fig.~\ref{Fig7b}(a,b) and for LRS device conductance in Fig.~\ref{Fig7b}(c,d) respectively. 
A clear separable distribution for each selected MLC state can be observed in Fig.~\ref{Fig7b}. HRS region states exhibit lesser current dissipation (compared to LRS region states) during READ operation for inference, leading to better energy efficiency. Accuracy results for simulated and ideal networks (full-precision and QNN) are summarized in Table \ref{tab1}. Compared to full-precision networks, the ideal BNN and TNN show an accuracy drop of $\sim$8\% and $\sim$6\% while offering memory savings of 32$\times$ and 16$\times$ respectively. In case of the proposed OxRAM-based BNN implementation, there is a marginal increase in the performance compared to ideal BNN that can be attributed to the imbalance resulting from synaptic weight mapping i.e. unlike symmetric `+1'/ `-1' weights, they are realized as positive conductance values relating to {LRS/}HRS distributions of the devices. For the proposed TNN, accuracy drop of 2\% is observed compared to an ideal TNN {in case of HRS based MLC state mapping and 4\% in case of LRS based MLC state mapping. This can be attributed to better separability (i.e. relative window between states `-1' and `0') in distributions exhibited for HRS MLC states as compared to LRS MLC states.} To illustrate impact on class-wise accuracy (for 10 classes in FMNIST), confusion matrices for BNN/TNN for HRS region state mapping are shown in Fig.~\ref{Fig8}(a,b) respectively. Similarly, confusion matrices for BNN/TNN for LRS region state mapping are shown in Fig.~\ref{Fig8}(c,d) respectively. It can be observed that similar classes such as (0, 2, 4, 6) with matching features exhibit more false positives compared to other classes. False positive count increases even further when using LRS MLC states with TNN which can be attributed to sensing region overlap as observed in Fig.~\ref{Fig6a}.

\section{Conclusion}
\label{conc}
Hybrid CMOS-OxRAM neuronal and synaptic building blocks for realizing low-resource BNN/ TNN have been experimentally demonstrated. Through extensive characterization, realization of reliable MLC OxRAM states with low C2C/D2D variability in order to store synaptic weights on a single 1T-1R bitcell was validated. A differential read scheme was presented for performing vector multiplication. BNNs/TNNs were simulated exploiting aforementioned MLC OxRAM states utilizing both LRS and HRS states. Simulated network showed performance degradation (2-5\%) compared to ideal TNN realized and a 0.18\% improvement compared to ideal software BNN highlighting impact of sensitivity to MLC state distributions. {While there may be marginal limitations in the proposed approach in terms of area, it still shows significant energy saving in scenarios where weight re-programming is performed ($\sim$2$\times$) at regular intervals.}

%\section*{Acknowledgements}
%The authors gratefully acknowledge CEA-LETI CMP MAD service for fabrication of samples used in this study.

\bibliographystyle{IEEEtran}
\bibliography{ref}

\end{document}